\date{}
\documentclass[aps,pra,showpacs]{revtex4-1}
\usepackage{graphicx,psfrag,amsmath,amssymb,amsfonts,latexsym,color,dcolumn}
\begin{document}
\title{Electrostatic Interaction due to Patch Potentials on Smooth
Conducting Surfaces}
\author{C\'esar D. Fosco$^{1,2}$}
\author{Fernando C. Lombardo$^3$}
\author{Francisco D. Mazzitelli$^{1,2}$}
\affiliation{$^1$ Centro At\'omico Bariloche,
Comisi\'on Nacional de Energ\'\i a At\'omica,
R8402AGP Bariloche, Argentina}
\affiliation{$^2$ Instituto Balseiro,
Universidad Nacional de Cuyo,
R8402AGP Bariloche, Argentina}
\affiliation{$^3$ Departamento de F\'\i sica {\it Juan Jos\'e
Giambiagi}, FCEyN UBA, Facultad de Ciencias Exactas y Naturales,
Ciudad Universitaria, Pabell\' on I, 1428 Buenos Aires, Argentina - IFIBA}
\date{today}
\begin{abstract}
 We evaluate the electrostatic interaction energy between two
surfaces, one flat and the other slightly curved, in terms of the two-point
autocorrelation functions for patch potentials on each one of them, and
of a single function $\psi$ which defines the curved surface.  The
resulting interaction energy, a functional of $\psi$, is evaluated up to
the second order in a derivative expansion approach.  We derive explicit
formulae for the coefficients of that expansion as simple integrals
involving the autocorrelation functions, and evaluate them for some
relevant patch-potential profiles and geometries.  
\end{abstract}
\maketitle 
\section{Introduction}\label{sec:intro} 
Mostly because of the ever improving precision achieved by Casimir force
experiments~\cite{books}, it has become important to be able to
discriminate that force from other interactions which, under some
circumstances, may even mask it completely.  Among them, the most relevant
ones are usually electrostatic in nature.  Indeed, in the usual
experimental setup one deals with metallic mirrors which, even if
electrically neutral,  may still produce relevant residual effects.
Surface imperfections can lead to a local departure from ideal metallic
behaviour, yielding space-dependent `patch' potentials on the mirrors'
surfaces. Those potentials, which may be interpreted as due to the
existence of superficial dipole layers on otherwise neutral
bodies~\cite{Lang}, can indeed produce a force. This force may be relevant
in precision experiments related to many different areas of
physics~\cite{experiments,dalvit13}.

For the case of two infinite parallel planes, that force has been studied
in detail by Speake and Trenkle in Ref.\cite{Speake}.  The force is,
naturally, a function of $a$, the distance between the mirrors, and of 
the patch potentials on the surfaces.  It has been shown, however, that the
force depends on the potentials only through their autocorrelation
functions. A conceptually different situation, which however leads to a
formally identical result corresponds to random potentials. In those cases,
the correlation functions is all the information we have about the
underlying potentials.

Recent Casimir effect experiments do involve, however, different
geometries. Of particular interest is the case of a sphere in front of a
plane; the influence of patch potentials in Casimir measurements for this
geometry has been recently analyzed in Ref.\cite{Behunin1}, using a
Proximity Force Approximation (PFA) \cite{Derjaguin}. The same system has
also been studied by using an exact numerical approach~\cite{Behunin2}.
The PFA approach amounts, in this context, to a local application of the
parallel plate result, and may therefore be implemented based in the
results of Ref.\cite{Speake}.

It is the aim of this paper to go beyond the PFA, by extending the results
for the patch potentials force, known for parallel planes, to a class of
different geometries, namely, to the case of an infinite plane facing a
slightly curved surface. By the latter we mean a surface which curvature
radius is much larger than the distance between the surfaces.  Moreover, we
know that in this situation the Casimir interaction may be calculated using
a Derivative Expansion (DE) approach~\cite{DE}, of which the Proximity
Force Approximation (PFA) is the leading term. We shall construct here the
same approximation for the patch potentials force, namely, the first
nontrivial correction to the PFA.

With the extra requirement that the autocorrelation lengths should also be
smaller than the curvature radius of the curved surface,  we shall be able
to apply here essentially the same approach used in the Casimir case to the
calculation of the leading and subleading terms for the patch potential
electrostatic interaction.  The leading term will of course amount to the
well-known PFA, while the subleading one shall account for the first
correction to it. 

To be more precise about the system that we consider, it will consist of
two static surfaces, denoted by $L$ and $R$, embedded in $3$-dimensional
space. $L$ shall be assumed to be flat, and, by a proper choice of
coordinates, to coincide with the $x_3=0$ plane ($x_1$, $x_2$ and $x_3$
denote the three Cartesian coordinates of a point ${\mathbf x}$ in this
coordinate system). The $R$ surface is, on the other hand, assumed to be
definable by a single function, namely, by an equation of the
form~\mbox{$x_3 = \psi({\mathbf x_\parallel})$}, where \mbox{${\mathbf
x_\parallel}=(x_1,x_2)$}.  This assumption does not cover all the possible
curved surfaces; however, it is sufficiently general as to describe the
usually considered experimental setups.  As advanced, the $R$ surface will
be assumed to be smooth, something that will be quantified below in terms
of $\psi$, its derivatives, and the correlation lengths characterizing the
patch potentials.  The extension to the case of two slightly deformed
surfaces is straightforward.

In a previous work~\cite{ann}, dealing with a DE approach to the
electrostatic interaction between perfect conductors, we analyzed what may
be regarded as a limiting case of a system with patch potentials. Namely,
one in which the surfaces $L$ and $R$ are held at constant potentials, so
that there is a constant electrostatic potential difference $V$ between
them.  There, at second order, the DE of the electrostatic energy
$U_{DE}$ stored between the surfaces was found to be:
\begin{equation}
U_{DE}=\frac{\epsilon_0 V^2}{2}\int d^2{\mathbf x_\parallel}\, 
\frac{1}{\psi}\left[ 1 +\frac{1}{3}(\nabla\psi)^2\right] \;.
\label{DE perf}
\end{equation} 
On the other hand (for a system with the same geometry) in the presence of
patch potentials, on general grounds we expect a DE expansion (up to the same
order) to have the form:
\begin{equation}
U_{DE}=\frac{\epsilon_0}{2}\int d^2{\mathbf x_\parallel}\, \left[ V(\psi) + Z(\psi)(\nabla\psi)^2\right]\, ,
\label{DE expected}
\end{equation} 
where  $V(\psi)$ and $Z(\psi)$ are functions that depend on the shape of
the $R$ surface and on the correlation functions for the patch potentials on
both surfaces. One of our main goals in this paper is to find an explicit
expression for the function $Z(\psi)$ (the zeroth-order, PFA contribution
has already been computed in previous works). 

This paper is organized as follows: in Section~\ref{sec:energy}, we obtain
a formal general expression for the electrostatic interaction energy
between the two surfaces, in terms of the correlation function between the
patch potentials, and of the exact Dirichlet Green's function for the
spatial region delimited by the two surfaces.  Based on that formal
expression we obtain, in Section~\ref{sec:expansion}, the electrostatic
interaction energy for small departures from the parallel planes case. In
Section~\ref{sec:deriv expansion}, we use those results to expand the
interaction energy up to the second order in derivatives, and apply the
results to different geometries and different models for the correlation
functions of the patch potentials. Our final remarks are included in Section
\ref{sec:conc}. The Appendix contain some details of the expansion of the
Green's function for almost flat surfaces.  
\section{Interaction energy}\label{sec:energy} 
In order to obtain the interaction energy, we calculate the electrostatic
potential $\varphi$, solution to an electrostatic problem in the region
${\mathcal V}$, limited by the surfaces $L$ and $R$, namely, $0 \leq x_3
\leq \psi({\mathbf x_\parallel})$.  A proper definition of the problem
requires some care, since one wants to consider the seemingly inconsistent
situation of having space dependent potentials on the surfaces of
conducting bodies. We bypass that problem by following the approach 
of Ref.\cite{Speake}, where the conflicting conditions are introduced on 
(infinitesimally close) different surfaces.  
More concretely, the electrostatic problem is defined for the
volume ${\mathcal V}$, with Dirichlet type boundary conditions on the $L$
and $R$ surfaces.
Patch potentials are then introduced not on $L$ and $R$, but on two
slightly shifted surfaces, $L_\delta$ and $R_\delta$, defined by $x_3 = 
\delta$ and $x_3 = \psi({\mathbf x_\parallel}) - \delta$ ($\delta >0$),
respectively. Potentials are not introduced as boundary conditions (they have already been
defined on $L$ are $R$); rather, one models them by means of two electric dipole
layers with surface densities ${\mathbf m_{L,R}}(\mathbf x_\parallel)=
m_{L,R}(\mathbf x_\parallel)\mathbf{\hat{n}}_{L,R}$,
($\mathbf{\hat{n}}_{L,R}$: unit normal to the respective surface). 

In the limit when the layers are adjacent to the surfaces, those moment
distributions generate variations in the surface potentials,  determined by
two functions,
\mbox{$\varphi_{L,R}({\mathbf x_\parallel})= m_{L,R}(\mathbf
x_\parallel)/\epsilon_0$}, which indeed do fix the
potential $\varphi$ on the inner face of $L$ and $R$, respectively. 

Thus, the electrostatic problem amounts to one where the charge density
inside ${\mathcal V}$ (associated to the dipole layers above) is given by:
\begin{equation}\label{rho}
\rho({\mathbf x}) = \epsilon_0 \Big[
\varphi_L({\mathbf x_\parallel})\, {\mathbf{\hat
n}}_L\cdot\nabla\delta(x_3-\delta) 
+ \varphi_R({\mathbf x_\parallel})\, {\mathbf{\hat
n}}_R\cdot\nabla\delta(x_3-\psi({\mathbf x_\parallel})+\delta) \Big] \;,
\end{equation}
where $\delta \to 0$, the limit whereby the dipole layers approach the two
surfaces `from inside' shall be taken when deriving the expression for the
energy (see below). 

The energy $U$ may be written in terms of the scalar potential
$\varphi({\mathbf x})$, as follows: 
\begin{equation}\label{eq:defw}
U\,=\,\frac{\epsilon_0}{2} \int_{\mathcal V} d^3{\mathbf x} \, ({\mathbf \nabla}
\varphi({\mathbf x}))^2 \;.  
\end{equation}
To proceed, we need to write $U$ in terms of $\varphi_L$ and $\varphi_R$.
The formal solution for $\varphi$ inside ${\mathcal V}$ may be written in
terms of the Dirichlet Green's function $G({\mathbf x},{\mathbf y})$,
defined for ${\mathbf x}$ and ${\mathbf y}$ in ${\mathcal V}$, as follows:
\begin{equation}
\varphi({\mathbf x}) \;=\;\frac{1}{\epsilon_0}\int_{{\mathbf x'} 
\in \mathcal V} G({\mathbf x},{\mathbf x'}) \rho({\mathbf x'})\, .
\label{solgen}
\end{equation}
In our conventions, $G$ is determined by the equations:
\begin{equation}\label{eq:green}
- \nabla^2_{\mathbf x} G({\mathbf x},{\mathbf y}) \;=\; \delta({\mathbf
  x}-{\mathbf y}) \;\;,\;\;\; G({\mathbf x},{\mathbf y}) \;=\; 0 \;,\;\;
\forall {\mathbf x}\in {\mathcal V} \;,\;\; \forall {\mathbf y} \in
{\mathcal S}= L \cup R \;. 
\end{equation} 
Besides, the Dirichlet Green's function is symmetric: $G({\mathbf
x},{\mathbf y})=G({\mathbf y},{\mathbf x})$.  

Inserting Eq.(\ref{rho}) into Eq.(\ref{solgen}), we obtain
\begin{equation}\label{eq:solphi} 
\varphi({\mathbf x}) \;=\; - \, 
\oint_{{\mathbf x'} \in \mathcal {\mathcal S}_\delta} d\sigma' \, {\varphi}({\mathbf x'}) \, 
\frac{\partial}{\partial n'} G({\mathbf x},{\mathbf x'}) 
\end{equation}
where ${\mathcal S_\delta} \equiv L_\delta \cup R_\delta$ denotes the
surfaces of the two layers, and  $\frac{\partial}{\partial n'}$ is the
directional derivative along the outer normal to ${\mathcal S_\delta}$ at each point on the surface, and
$d\sigma'$ is the surface element at each point ${\mathbf x'} \in {\mathcal
S}_\delta$. Note that the r.h.s. of the equation above is
completely determined by the prescribed values ${\varphi}_{L,R}$ of the
patch potentials on $L_\delta$ and $R_\delta$.

Inserting now (\ref{eq:solphi}) into (\ref{eq:defw}), integrating by parts, 
using (\ref{eq:green}), and then letting $\delta \to 0$, we see that $U$ may be written as follows:
\begin{equation}\label{eq:wg} 
U \;=\; \frac{\epsilon_0}{2} \, \oint_{\mathcal S}
d\sigma' \oint_{\mathcal S} d\sigma'' \; \varphi({\mathbf x'}) \,{\mathcal
K}({\mathbf x'},{\mathbf x''}) \,\varphi({\mathbf x''}) \;, 
\end{equation}
where we have removed the $\delta$ subscript for the surfaces since the
$\delta \to 0$ has been taken. The kernel ${\mathcal K}$ is constructed in terms of the normal
derivatives of the exact Green's function $G$ on ${\mathcal S}$:
\begin{equation}\label{eq:defk} 
{\mathcal K}({\mathbf x'},{\mathbf
x''})\,=\, \partial_{n'} \partial_{n''} G({\mathbf x'},{\mathbf x''}) \;,
\;\; {\mathbf x'},\, {\mathbf x''} \in {\mathcal S} \;.  
\end{equation} 
Note that the presence of the metallic media has been taken into account,
by discarding boundary terms which involve the values of the Green's
function on those media. At this point the distinction between ${\mathcal
S}$ and ${\mathcal S}_\delta$ ceases to be necessary, since we have ended
up with an expression where one has to integrate on the inner faces of the
mirrors only. 

Expression (\ref{eq:wg}) has to be now particularized to the case
of the region we are considering, and the assumed values for the potential
on the boundary. Denoting by ${\mathcal K}_{AB}$, $A, B = L, R$ the special
form adopted by the kernel ${\mathcal K}$ when its first argument belong to
the surface $A$ and its second argument to $B$, we see that:
\begin{equation} 
U \;=\; \sum_{A,B} \, U_{AB} 
\end{equation} where:
\begin{eqnarray} 
U_{LL} &=& \frac{\epsilon_0}{2} \, \int d^2{\mathbf
x_\parallel'}\int d^2{\mathbf x_\parallel''} \; \varphi_L({\mathbf
x_\parallel'}) \,{\mathcal K}_{LL} ({\mathbf x_\parallel'},{\mathbf
x_\parallel''}) \,\varphi_L({\mathbf x_\parallel''}) \nonumber\\ 
U_{LR} &=& \frac{\epsilon_0}{2}
\, \int d^2{\mathbf x_\parallel'}\int d^2{\mathbf
x_\parallel''} \sqrt{g({\mathbf x_\parallel''})} \; \varphi_L({\mathbf
x_\parallel'}) \,{\mathcal K}_{LR}({\mathbf x_\parallel'},{\mathbf
x_\parallel''}) \,\varphi_R({\mathbf x_\parallel''})\nonumber\\ 
U_{RL} &=& \frac{\epsilon_0}{2} \, \int d^2{\mathbf x_\parallel'}\int d^2{\mathbf
x_\parallel''}\, \sqrt{g({\mathbf x_\parallel'})} \, \varphi_R({\mathbf
x_\parallel'}) \,{\mathcal K}_{RL}({\mathbf x_\parallel'},{\mathbf
x_\parallel''}) \,\varphi_L({\mathbf x_\parallel''}) \nonumber\\ 
U_{RR} &=&  \frac{\epsilon_0}{2} \, \int d^2{\mathbf x_\parallel'} \int d^2{\mathbf
x_\parallel''} \sqrt{g({\mathbf x_\parallel'})} \sqrt{g({\mathbf
x_\parallel''})} \, \varphi_R({\mathbf x_\parallel'}) \,{\mathcal
K}_{RR}({\mathbf x_\parallel'},{\mathbf x_\parallel''})
\,\varphi_R({\mathbf x_\parallel''}) \;, 
\end{eqnarray} 
where we have written the area element in terms of the parametrization of
the respective surface, namely, on $L$, $d^2\sigma = d^2{\mathbf
x_\parallel}$, while on $R$: $d^2\sigma = d^2{\mathbf x_\parallel} \sqrt{1
+ (\nabla \psi({\mathbf x_\parallel}))^2}$.   

The explicit form of the outer normal derivative depends on the surface,
$L$ or $R$, considered. For $L$, we have $\partial_n = - \partial_3$ while
for $R$: 
\begin{equation} 
\partial_n = \frac{\partial_3 - (\partial_i \psi)
\partial_i}{\sqrt{1 + (\nabla \psi({\mathbf x_\parallel}))^2}} \;,
\end{equation} 
where $i$ and, in what follows, indices from the middle of
the Roman alphabet $(i,\,j,\,\ldots)$ run from $1$ to $2$.

We may write then a more explicit form for $U_{AB}$, as follows:

\begin{eqnarray}\label{eq:resf} 
U_{LL} &=& \frac{\epsilon_0}{2} \, \int_{{\mathbf
x_\parallel'},{\mathbf x_\parallel''}} \; \big[\partial'_3 \partial''_3
G({\mathbf x'},{\mathbf x''})\big]_{x'_3=x''_3=0} \varphi_L({\mathbf
x_\parallel''}) \varphi_L({\mathbf x_\parallel'}) \nonumber\\ 
U_{LR} &=& -  \frac{\epsilon_0}{2} \, \int_{{\mathbf x_\parallel'},{\mathbf x_\parallel''}}
\big[\partial'_3 (\partial''_3  - \partial''_i\psi \partial''_i) G({\mathbf
x'},{\mathbf x''}) \big]_{x'_3=0, x''_3=\psi({\mathbf x_\parallel''})}
\varphi_R({\mathbf x_\parallel''}) \varphi_L({\mathbf x_\parallel'})
\nonumber\\ 
U_{RL} &=& - \frac{\epsilon_0}{2} \, \int_{{\mathbf
x_\parallel'},{\mathbf x_\parallel''}} \big[(\partial'_3  - \partial'_i\psi
\partial'_i) \partial''_3 G({\mathbf x'},{\mathbf x''})
\big]_{x'_3=\psi({\mathbf x_\parallel'}), x''_3=0} \varphi_L({\mathbf
x_\parallel''}) \varphi_R({\mathbf x_\parallel'}) \nonumber\\ 
U_{RR} &=& \frac{\epsilon_0}{2} \, \int_{{\mathbf x_\parallel'},{\mathbf x_\parallel''}}
\big[(\partial'_3  - \partial'_i\psi \partial'_i)(\partial''_3  -
\partial''_j\psi \partial''_j) G({\mathbf x'},{\mathbf x''})
\big]_{x'_3=\psi({\mathbf x_\parallel'}), x''_3 \in \psi({\mathbf
x_\parallel''})} \varphi_R({\mathbf x_\parallel''}) \varphi_R({\mathbf
x_\parallel'}) \;, 
\end{eqnarray} 
where the $\sqrt{1 + (\nabla \psi({\mathbf x_\parallel}))^2}$ factors from the surface element on $R$
have been exactly cancelled with identical ones in the denominators of the normal derivatives.
In order to simplify the expressions, in the equation above and in what follows we omit the integration 
measure and use the notation
\begin{equation}
 \int d^2{\mathbf x_\parallel'}\int d^2{\mathbf
x_\parallel''}\rightarrow  \int_{{\mathbf
x_\parallel'},{\mathbf x_\parallel''}} \,\, \; .
\end{equation}
 
\section{Perturbative expansion}\label{sec:expansion} 

To obtain the DE, following the approach we introduced in
previous references, we consider now the situation \mbox{$\psi({\mathbf
x_\parallel}) = a + \eta({\mathbf x_\parallel})$} with $|\eta({\mathbf
x_\parallel})|<< a$, and expand $U$ in powers of $\eta$, up to the second
order. 

This expansion shall induce the corresponding expansion for $U$,
\begin{equation} 
U \;=\; U^{(0)} \,+\, U^{(1)}\,+\, U^{(2)} \,+\,\ldots
\end{equation} 
where the index denotes the corresponding order in $\eta$.
Also, each order can be further decomposed, as follows: 
\begin{equation}
U^{(\alpha)} \;=\; \sum_{A,B} \, U^{(\alpha)}_{A,B} \;.  
\end{equation}

The dependence on $\eta$ in each term of the sum above will come from the
Green's function dependence on $\psi$, some details of which are presented
in the Appendix.  This shall produce different contributions to each order,
which we will consider now in turn.  For the $\eta^{0}$ order, we have:
\begin{equation} 
U^{(0)}_{LL} \,=\,    \frac{\epsilon_0}{2} \, \int_{{\mathbf
x_\parallel'},{\mathbf x_\parallel''}} \; \big[\partial'_3 \partial''_3
G^{(0)}({\mathbf x'},{\mathbf x''})\big]_{x'_3=0, x''_3=0}
\varphi_L({\mathbf x_\parallel''}) \varphi_L({\mathbf x_\parallel'}) \;,
\end{equation} 
\begin{equation} 
U^{(0)}_{LR} \;=\; - \frac{\epsilon_0}{2} \,
\int_{{\mathbf x_\parallel'},{\mathbf x_\parallel''}} \big[\partial'_3
\partial''_3 G^{(0)}({\mathbf x'},{\mathbf x''}) \big]_{x'_3=0, x''_3=a}
\varphi_R({\mathbf x_\parallel''}) \varphi_L({\mathbf x_\parallel'}) \;,
\end{equation} 
\begin{equation} U^{(0)}_{RL} \;=\; - \frac{\epsilon_0}{2} \,
\int_{{\mathbf x_\parallel'},{\mathbf x_\parallel''}} \big[\partial'_3
\partial''_3 G^{(0)}({\mathbf x'},{\mathbf x''}) \big]_{x'_3=a, x''_3=0}
\varphi_L({\mathbf x_\parallel''}) \varphi_R({\mathbf x_\parallel'})
\end{equation}
and 
\begin{equation} 
U^{(0)}_{RR} \;=\; \frac{\epsilon_0}{2} \,
\int_{{\mathbf x_\parallel'},{\mathbf x_\parallel''}} \big[\partial'_3
\partial''_3 G^{(0)}({\mathbf x'},{\mathbf x''}) \big]_{x'_3=a, x''_3=a}
\varphi_R({\mathbf x_\parallel''}) \varphi_R( {\mathbf x_\parallel'}) \;.
\end{equation}

It is now convenient to note that, since the kernels depending on the
$0^{th}$ order Green's function are translation invariant along the
parallel directions, they can only depend on the difference ${\mathbf
x'_\parallel}-{\mathbf x''_\parallel}$. Thus, under a shift of arguments in
the integrals, and defining ${\mathbf y_\parallel} \equiv {\mathbf
x_\parallel''}-{\mathbf x_\parallel'}$, it is straightforward to see that
the respective energies per unit area, ${\mathcal U}^{(0)}_{AB} \equiv
U^{(0)}_{AB}/{\mathcal A}$ (where ${\mathcal A}$ is the area of the flat
plate), can be written as follows: 
\begin{equation} 
{\mathcal U}^{(0)}_{LL} \,=\,     \frac{\epsilon_0}{2}\, \int_{{\mathbf y_\parallel}} \; 
\big[\partial'_3 \partial''_3 G^{(0)}({\mathbf x'},{\mathbf x''})\big]_{x'_3=0, x''_3=0}
\Omega_{LL}({\mathbf y_\parallel}) \;, 
\end{equation} 
\begin{equation}
{\mathcal U}^{(0)}_{LR} \;=\; - \frac{\epsilon_0}{2} \, \int_{\mathbf
y_\parallel} \big[\partial'_3 \partial''_3 G^{(0)}({\mathbf x'},{\mathbf
x''}) \big]_{x'_3=0, x''_3=a} \Omega_{RL}({\mathbf y_\parallel}) \;,
\end{equation} 
\begin{equation} 
{\mathcal U}^{(0)}_{RL} \;=\; - \frac{\epsilon_0}{2}  \, \int_{\mathbf y_\parallel} \big[\partial'_3 \partial''_3
G^{(0)}({\mathbf x'},{\mathbf x''}) \big]_{x'_3=a, x''_3=0}
\Omega_{LR}({\mathbf y_\parallel}) 
\end{equation} 
\begin{equation}
{\mathcal U}^{(0)}_{RR} \;=\; \frac{\epsilon_0}{2} \, \int_{{\mathbf
x_\parallel'},{\mathbf x_\parallel''}} \big[\partial'_3 \partial''_3
G^{(0)}({\mathbf x'},{\mathbf x''}) \big]_{x'_3=a, x''_3=a}
\Omega_{RR}({\mathbf y_\parallel}) \;, 
\end{equation} 
where $\Omega_{AB}$ denotes the correlation functions for the patch potentials, defined by:
\begin{equation} 
\Omega_{AB}({\mathbf x_\parallel}) \,=\, \frac{1}{\mathcal
A} \int d^2{\mathbf z_\parallel} \; \varphi_A({\mathbf x_\parallel} +
{\mathbf z_\parallel} ) \varphi_B({\mathbf z_\parallel}) \;.
\end{equation} 
Note that, because of the translation invariance, these contributions to
the energy depends on the correlations, and not on the potentials themselves.

The leading term of the interaction energy  can be obtained by evaluating the densities above in Fourier 
space. The result is
\begin{eqnarray} 
{\mathcal U}^{(0)} &=& - \epsilon_0 \int \frac{d^2{\mathbf
k_\parallel}}{(2\pi)^2} \, \frac{|{\mathbf k_\parallel}|}{e^{2|{\mathbf
k_\parallel}| a} - 1} \, \left( {\widetilde
\Omega}_{LL}({\mathbf k_\parallel}) + {\widetilde\Omega}_{RR}({\mathbf k_\parallel})\right)
\nonumber\\ &+& \epsilon_0   \int
\frac{d^2{\mathbf k_\parallel}}{(2\pi)^2} \, \frac{|{\mathbf
k_\parallel}|}{\sinh(|{\mathbf k_\parallel}|a) }
\, {\widetilde\Omega}_{LR}({\mathbf k_\parallel})\;, 
\label{eq:orden0}
\end{eqnarray} 
where we have used the symmetry ${\widetilde\Omega}_{LR}=
{\widetilde\Omega}_{RL}$; besides, ${\widetilde\Omega}_{AB}$ is by
construction a function of ${\mathbf k_\parallel}$.  It is worth
mentioning at this point that an identical result  would follow from the
assumption of the existence of purely statistical correlations between the
potentials.

The result of  Eq.(\ref{eq:orden0}) coincides, up to terms independent of
$a$, with that of Ref. \cite{Speake}. One could try to recover the usual
result for perfect conductors assuming that $\varphi_L =V$ and
$\varphi_R=0$.  In this case Eq.(\ref{eq:orden0}) gives ${\mathcal
U}^{(0)}=-\epsilon_0V^2/(2 a)$, i.e.  the usual result for a capacitor with
the opposite sign. This sign comes from the fact that we are computing the
electrostatic energy for two grounded surfaces with  dipole layers close to
them, that we are modelling with the potentials $\varphi_A$. The case of
two surfaces at different potentials has an additional surface term
(discarded in the formal derivation of the previous section) that would reverse the sign in the final answer. 
    
Regarding the $\eta^{1}$ order, it is quite straightforward to see that it
vanishes under the assumption that the spatial average of $\eta$ is zero.
This will be enough for our purpose of computing the DE (see below).

Finally, let us consider now the second order contributions. We shall
neglect here the $\Omega_{A\neq B}$ correlations, since they are assumed to
be strongly suppressed with the distance between plates (at least in
comparison with the autocorrelations $\Omega_{A=A}$). Thus we just need to
compute: 
\begin{eqnarray}\label{eq:o2} 
U^{(2)}_{LL} &=& \frac{\epsilon_0}{2} \,
\int_{{\mathbf x_\parallel'},{\mathbf x_\parallel''}} \; \big[\partial'_3
\partial''_3 G({\mathbf x'},{\mathbf x''})\big]_{x'_3=x''_3=0}^{(2)}
\varphi_L({\mathbf x_\parallel''}) \varphi_L({\mathbf x_\parallel'})
\nonumber\\ U^{(2)}_{RR} &=& \frac{\epsilon_0}{2} \, \int_{{\mathbf
x_\parallel'},{\mathbf x_\parallel''}} \; \big[\partial'_3 \partial''_3
G({\mathbf x'},{\mathbf x''})\big]_{x'_3=\psi({\mathbf x_\parallel'}),
x''_3=\psi({\mathbf x_\parallel''})}^{(2)}\, \varphi_R({\mathbf
x_\parallel''}) \varphi_R({\mathbf x_\parallel'}) 
\end{eqnarray}
which depend on two second-order objects, determined by the Green's functions.

To calculate this contribution we shall further assume that even when one
of the surfaces is curved, the product of the two potentials is appreciably
different from zero only when the distance between the two arguments is
small.
Assuming that this autocorrelation length is much smaller than the
curvature radius, we shall still assume the translation invariance of the
arguments in the Green's function on that region. In this way, those
contributions may be again written in terms of the autocorrelation
functions, namely, 
\begin{eqnarray}\label{eq:oo2} U^{(2)}_{LL} &=&
\frac{\epsilon_0}{2} \, \int_{{\mathbf x_\parallel'},{\mathbf x_\parallel''}} \;
\big[\partial'_3 \partial''_3 G({\mathbf x'},{\mathbf
x''})\big]_{x'_3=x''_3=0}^{(2)} \Omega_{LL}({\mathbf
x_\parallel''}-{\mathbf x_\parallel'}) \nonumber\\ U^{(2)}_{RR} &=&
\frac{\epsilon_0}{2}\, \int_{{\mathbf x_\parallel'},{\mathbf x_\parallel''}} \;
\big[\partial'_3 \partial''_3 G({\mathbf x'},{\mathbf
x''})\big]_{x'_3=\psi({\mathbf x_\parallel'}), x''_3=\psi({\mathbf
x_\parallel''})}^{(2)}\, \Omega_{RR}({\mathbf x_\parallel''}-{\mathbf
x_\parallel'})\;.  
\end{eqnarray}

Then, performing the expansion of the Green's function, a rather lengthy
but otherwise standard calculation shows that, for $A=L,R$,
\begin{equation}\label{eq:Ell2} 
U^{(2)}_{AA}  \;=\;\frac{1}{2} \,\int
\frac{d^2{\mathbf k_\parallel}}{(2\pi)^2} \, \big|{\tilde\eta}({\mathbf
k_\parallel})\big|^2 \, f_{AA} ({\mathbf k_\parallel}) \;, 
\end{equation}
with: 
\begin{equation}\label{eq:fll} 
f _{AA}({\mathbf k_\parallel}) =
-\epsilon_0 \int \frac{d^2{\mathbf p_\parallel}}{(2\pi)^2} \,
{\widetilde\Omega}_{AA}({\mathbf p_\parallel}) |{\mathbf p_\parallel}|^2
|{\mathbf p_\parallel}+{\mathbf k_\parallel}|
\frac{\coth(|{\mathbf p_\parallel}+{\mathbf k_\parallel}|a)}{\sinh^2(|{\mathbf p_\parallel}|a)}\, .
\end{equation} 
It is reassuring to verify that, as it happened with the zeroth-order term,
the perfect conductor limit is correctly obtained; namely, when  $\varphi_L
=V$ and $\varphi_R=0$, one has ${\widetilde\Omega}_{RR}=0$, 
${\widetilde\Omega}_{LL}=(2\pi)^2V^2\delta({\mathbf p_\parallel})$ and
Eq. (\ref{eq:Ell2}) reproduces, up to the sign,  the
result of Ref.~\cite{ann}. 

Eq. (\ref{eq:fll}) is the main result of this section.  Note that it will
not only allow us to compute the derivative expansion of the interaction
energy for a smooth $R$ surface, it could also be applied, in a more direct
fashion, to the calculation of the energy and force between almost flat
surfaces in an expansion in powers of the amplitude of the deformations.

\section{Derivative expansion}\label{sec:deriv expansion}

To determine the DE we need now to obtain explicit forms for the functions $V$
and $Z$. They can be read from the perturbative expansion of the previous
Section in a straightforward way, by using an argument~\cite{DE} which 
we briefly review here: the electrostatic energy can be thought as a functional  $U[\psi]$. Assuming
the existence of a local expansion for that functional, dimensional analysis implies that the first two  terms in the expansion
must have the form presented in Eq.(\ref{DE expected}).  Then, in order to determine the explicit form of $V$ and $Z$ it is
sufficient to evaluate $U[a + \eta({\mathbf x_\parallel})]$ with $|\eta({\mathbf
x_\parallel})|<< a$, expanding the result up to the second order in $\eta$. The leading term 
in this expansion determines $V(a)$, while the one proportional to $(\nabla\eta)^2$ fixes 
$Z(a)$. Note that it is not necessary to keep terms that are linear in $\eta$, nor terms proportional
to $\eta^2$ without derivatives.  The assumption of locality that leads to Eq.(\ref{DE expected}) can 
be formally analyzed as described in Section V of Ref.\cite{fourth}.

Using these arguments, the leading term of the DE can be
obtained by replacing $a$ by $\psi({\mathbf x_\parallel})$ in Eq.(\ref{eq:orden0}), and integrating over ${\mathbf x}_\parallel$: 
\begin{equation} 
U^{(0)}_{DE} = - \epsilon_0  \int d^2{\mathbf x_\parallel}\, \int \frac{d^2{\mathbf
k_\parallel}}{(2\pi)^2} \, \frac{|{\mathbf k_\parallel}|}{e^{2|{\mathbf
k_\parallel}| \psi({\mathbf x_\parallel})} - 1} \,\left( {\widetilde
\Omega}_{LL}({\mathbf k_\parallel}) + {\widetilde\Omega}_{RR}({\mathbf k_\parallel})\right)
\, ,
\end{equation} 
where, as before, we neglected the correlations between different surfaces.
Using the notation of Eq.(\ref{DE expected}),
and assuming isotropy for the correlation function we find
\begin{equation} 
V(\psi)= -  \frac{1}{\pi\psi^3}\int_0^\infty dx \frac{x^2}{e^{2x}-1} \,\left( {\widetilde
\Omega}_{LL}(x/\psi) + {\widetilde\Omega}_{RR}(x/\psi)\right)
\, .
\end{equation} 
 
To obtain the second order term, we expand the form factor $f_{AA}$ in powers of ${\mathbf k_\parallel}$
and keep the second order term, that is
\begin{eqnarray}
f _{AA}^{(2)}({\mathbf k_\parallel}) &=& \frac{\epsilon_0\vert{\mathbf k_\parallel}\vert^2}{32\pi a^3}\int_0^\infty
 dx\, \frac{x^2 \,{\widetilde\Omega}_{AA}(x/a)}{\sinh^5(x)}\times\nonumber\\
 &&\left [(1-8x^2)\cosh(x)-\cosh(3 x)+12 x\sinh(x)\right]\, .
 \end{eqnarray}
 Then we insert this expression into Eq.(\ref{eq:Ell2}) and replace
$a\to\psi$ and $\nabla\eta\to\nabla\psi$. This produces the result
 \begin{equation}
 U^{(2)}_{DE}=\frac{\epsilon_0}{2}  \int d^2{\mathbf x_\parallel}\,Z(\psi)\left(\nabla\psi\right)^2\, ,
 \label{u2}
 \end{equation}
 where
 \begin{eqnarray}
 Z(\psi)=\frac{1}{32\pi\psi^3} 
&& \int_0^\infty
 dx\, \frac{x^2 \left({\widetilde\Omega}_{LL}(x/\psi)+ {\widetilde\Omega}_{RR}(x/\psi) \right)  }{\sinh^5(x)}\times
 \nonumber\\ &&
\left [(1-8x^2)\cosh(x)-\cosh(3 x)+12 x\sinh(x)\right]\, .
\end{eqnarray}

Thus we have obtained the two expression that determine the second order
derivative expansion; they require the calculation of two integrals, 
involving the autocorrelation functions. 

\subsection{Evaluation for some particular models}

We now evaluate the functions $V(\psi)$ and $Z(\psi)$ for some particular
correlation functions. We will assume they are identical on both mirrors,
and use the notation ${\widetilde\Omega} \equiv {\widetilde\Omega}_{RR}
={\widetilde\Omega}_{LL}$ for their common profile.

Let us first assume that the correlation function depends on the variance of
the potential $V_{rms}$ and on a single characteristic length ${\ell }$. Then, on
dimensional grounds we shall have 
\begin{equation}
{\widetilde\Omega}({\mathbf k_\parallel})=V^2_{rms}\,{\ell}^2\,g(\vert{\mathbf k_\parallel}\vert{\ell}),
\end{equation}
for some dimensionless function of a dimensionless argument, $g$. 

It then follows that
\begin{equation}
V(\psi)=\frac{V^2_{rms}}{\psi}\,v({\ell}/\psi)\,\, \quad Z(\psi)=\frac{V^2_{rms}}{\psi}\, z({\ell}/\psi)\, ,
\end{equation}
where
\begin{eqnarray}
v({\ell}/\psi)&=&-  \frac{2}{\pi}\frac{\ell^2}{\psi^2}\int_0^\infty dx \frac{x^2}{e^{2x}-1}g(x\ell/\psi) \nonumber\\
z({\ell}/\psi)&=&\frac{1}{16\pi}\frac{\ell^2}{\psi^2} 
\int_0^\infty
 dx\, \frac{x^2 g(x\ell/\psi) }{\sinh^5(x)}\times
 \nonumber\\ &&
\left [(1-8x^2)\cosh(x)-\cosh(3 x)+12 x\sinh(x)\right]\, .
\end{eqnarray}
Assuming that the function $g$ tends to zero for large values of its argument, one can show that $v$ and $z$ tend to the result for constant potentials
when $\ell\ll \psi$. To see this, we note that in this limit the main contribution to the integrals come from $x\ll 1$. Therefore, we can expand 
the functions that multiply $g$ in the integrals around $x=0$ to obtain
\begin{equation}
v \simeq 3z\simeq-\frac{1}{\pi}\int_0^\infty dy\, y \, g(y)=-2\, ,
\label{largep}
\end{equation}
where we have used the general property \cite{Behunin1}
\begin{equation}
V_{rms}^2=\int\frac{d^2{\mathbf k}_\parallel}{(2\pi)^2}\, {\widetilde \Omega}({\mathbf k}_\parallel)\, .
\end{equation}
Note that Eq.(\ref{largep}) correspond to twice the constant potential result, because we are considering the same correlation function
on both surfaces.

On the other hand, in the opposite limit $\ell\ll \psi$, one can make the approximation  $g(x\psi/\ell)\simeq g(0)$ inside the integrals to get
\begin{equation}
v\simeq -\frac{g(0)\zeta(3)}{2\pi} \frac{\ell^2}{\psi^2}\quad z\simeq -\frac{g(0)(1+6\zeta(3))}{24\pi} \frac{\ell^2}{\psi^2}.\label{smallarg}
\end{equation} 
Notably, the interaction energy for small patch potentials has the same 
dependence with distance as the Casimir energy. 

\subsection{Quasilocal correlations}

As an example of this class of correlation functions, let us consider the gaussian approximation to the quasilocal correlation function proposed in Ref.\cite{dalvit13}:
\begin{equation}
\Omega({\mathbf x}_\parallel)= V_{rms}^2 \exp[-4\vert  {\mathbf x}_\parallel\vert^2/\ell^2 ]\Rightarrow {\widetilde \Omega}({\mathbf k}_\parallel)=
\frac{\pi}{4}V_{rms}^2\ell^2 \exp[-\frac{1}{16}\vert  {\mathbf
k}_\parallel\vert^2 \ell^2 ] \;.
\end{equation}

\begin{figure}
\centering
\includegraphics[width=10cm , angle=0]{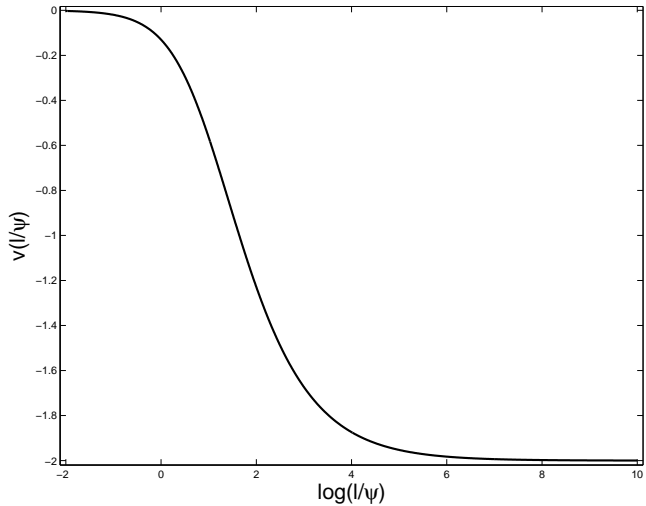}
\caption{$v(l/\psi$) as a function of the dimensionless patch correlation length $l/\psi$. We consider the Gaussian approximation to the 
quasilocal correlation function. It is possible to check that $v$ approaches to the 
correct value given by the Eq.(\ref{smallarg}) for $l/\psi \ll 1$, and it approaches the predicted value $v = -2$ for larger arguments. } \label{fig1}
\end{figure}

The functions $v$ and $z$ can be computed numerically; the results are
shown in Figs.1 and 2. As expected, the numerical results have the correct behavior in the limits of small and large correlation lengths described in Eqs.(\ref{largep}) and
(\ref{smallarg}) with $g(0)=\pi/4$.

\begin{figure}
\centering
\includegraphics[width=10cm , angle=0]{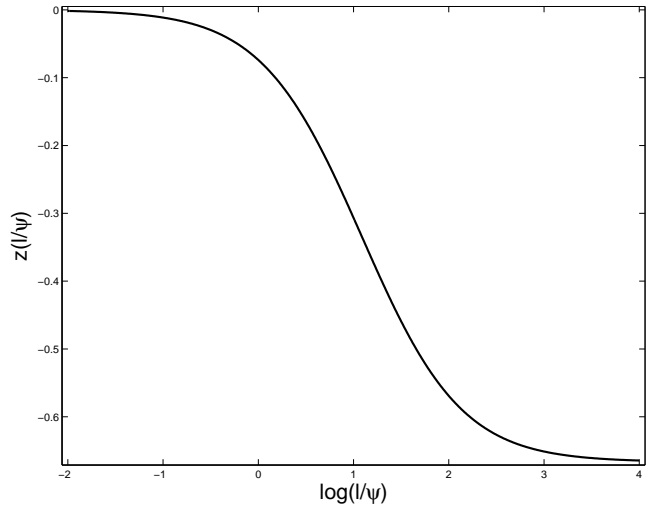}
\caption{$z(l/\psi$) as a function of the dimensionless patch correlation length $l/\psi$ in the Gaussian model. It is possible to check that it approaches to the 
correct value given by the Eq.(\ref{smallarg}) for $l/\psi \ll 1$ and also to the predicted value $z = -\frac{2}{3}$ when $l \gg \psi$.} \label{fig2}
\end{figure}

We also show the rate between the second and zeroth order contributions to the electrostatic energy, for the case of having a 
sphere in front of a plane. We consider a sphere of radius $R$ at a distance $a$ from a plane. Denoting 
by ($r$,$\phi$) the polar coordinates in the ($x_1$,$x_2$) plane the function $\psi$ 
reads

\begin{equation}\psi (r) = R   \left[ \frac{a}{R} + 1 - \sqrt{1 - \frac{r^2}{R^2}}\right], 
\end{equation}
In the Fig. 3 we plot the ratio ${\cal Q} = \frac{\left[ U^{(0)}_{DE} + U^{(2)}_{DE} \right]}{ U^{(0)}_{DE}}$ as a function of $\epsilon = a/R$. As expected, there is 
a linear correction for small values of $a/R$.  As the dependence with $\ell$ is the same for $U^{(0)}_{DE}$ and $U^{(2)}_{DE}$ in the small patches limit $\ell\ll a$ (Eq.\ref{smallarg}),
the value of ${\cal Q}$ does not depend on $\ell$ for $\ell/R\ll a/R$. We have checked this numerically.

\begin{figure}
\centering
\includegraphics[width=10cm , angle=0]{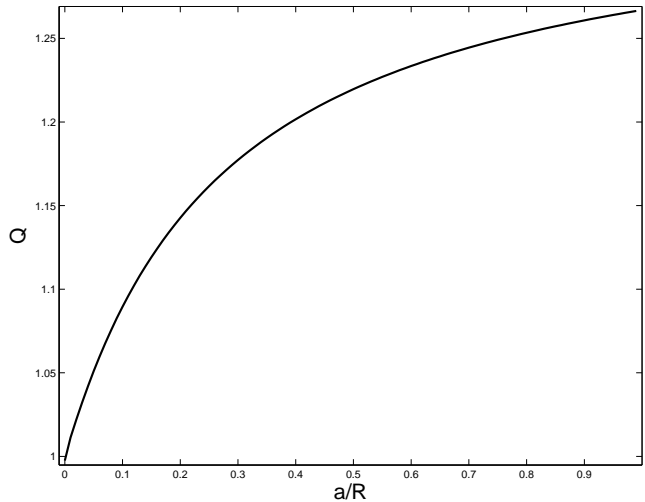}
\caption{Rate ${\cal Q}$ as a function of $a/R$ in the Gaussian approximation to the quasilocal patch correlation function, 
for $l/R = 0.001$. There is a lineal correction to the zeroth order.} \label{fig3}
\end{figure}

\subsection{Correlation function with sharp-cutoffs }
 
 As a second example, we will consider the sharp-cutoff model proposed in Ref.\cite{Speake}:
\begin{equation}
{\widetilde\Omega}({\mathbf k}_\parallel)= \frac{4\pi V^2_{rms}}{k_{max}^2-k_{min}^2}\theta(\vert{\mathbf k}_\parallel\vert-k_{min})\theta(k_{max}-\vert{\mathbf k}_\parallel\vert)
\label{VZ}
\end{equation}
 
 For this particular correlation, the functions $V$ and $Z$ can be computed analytically
\begin{eqnarray}
V(\psi)&=&-\frac{8 V^2_{rms}}{(k_{max}^2-k_{min}^2)\psi^3}\left[f_0(k_{max}\psi)-f_0(k_{min}\psi)\right]\nonumber\\
Z(\psi)&=&\frac{ V^2_{rms}}{4(k_{max}^2-k_{min}^2)\psi^3}\left[f_2(k_{max}\psi)-f_2(k_{min}\psi)\right]\, ,
\label{VZsharp}
\end{eqnarray}
where the functions $f_0$ and $f_2$ can be written in terms of polylogarithm functions $Li_n(z)$, and are given by
\begin{eqnarray}
f_0(x)&=&\frac{1}{2}x^2\log(1-e^{-2x})-\frac{1}{2} x\, Li_2(e^{-2x})-\frac{1}{4}  Li_3(e^{-2x})\nonumber\\
f_2(x)&=&\frac{1}{3} (-8 x^3+8 x^3 \coth(x)-4 x^3 \frac{\cosh(x)}{ \sinh(x)^3}+6 \frac{x^4}{ \sinh(x)^4}\nonumber\\&-&24 x^2 \log(1-e^{-2 x})
+24 x Li_2(e^{-2 x})+12 Li_3(e^{-2 x})
\, .
\label{poly}
\end{eqnarray}
These expressions contain as a particular case the constant potentials result, which can be obtained setting $k_{min}=0$ and then taking the limit 
$k_{max}=0$ (note that in this case the correlation function $\widetilde\Omega$ becomes a $\delta$-function).  Indeed, expanding $f_0$ and $f_2$
around $x=0$ one can show that $V(\psi)$ and $3 Z(\psi)$ tend to $ -2V_{rms}^2/\psi$. 

In the opposite limit, $k_{min}=0$, $k_{max}\to\infty$ we obtain
\begin{equation}
V\simeq -2\zeta(3)\frac{V_{rms}^2}{k_{max}^2\psi^3}\quad Z\simeq -\frac{1+6\zeta(3)}{6} \frac{V_{rms}^2}{k_{max}^2\psi^3}\, ,
\label{omegaconst}
\end{equation}
which, as expected,  coincide with the short correlation limit given in Eq.(\ref{smallarg}).

An important feature of the sharp-cutoff model is that the electrostatic interaction energy between the surfaces is strongly suppressed when
$k_{min}\gg 1/\psi$.  In this limit one can show that both $V$ and $Z$ are proportional to $\exp(-2k_{min}\psi)/\psi$.

\section{Conclusions}\label{sec:conc}  
We have found explicit results for the leading and subleading
terms in a DE approach to the calculation of the force due to patch
potentials, for a family of experimentally relevant geometric
configurations, in terms of the path potential correlation functions (which
appear in our results inside simple integrals). Regarding those correlation
functions, we have used specific models to obtain more explicit expressions
for the coefficient functions the determine the DE to the interaction
energy.

For the case of a gaussian correlation function, the DE of the interaction energy
interpolates between (twice) the electrostatic result for a constant differential potential 
for large patches ($\ell\gg\psi $)
\begin{equation}
U_{DE}=-\epsilon_0 V_{rms}^2\int d^2{\mathbf x_\parallel}\, 
\frac{1}{\psi}\left[ 1 +\frac{1}{3}(\nabla\psi)^2\right] \; ,
\end{equation}
and a Casimir-like interaction energy for small patches ($\ell\ll\psi$)
\begin{equation}
U_{DE}=-\frac{\epsilon_0 V_{rms}^2\ell^2\zeta(3)}{16}\int d^2{\mathbf x_\parallel}\, 
\frac{1}{\psi^3}\left[ 1 +0.569(\nabla\psi)^2\right] \;.
\end{equation}
Similar results are valid  for other correlation models, as the sharp-cutoff model,
unless the correlation function ${\widetilde\Omega}({\mathbf k}_\parallel)$  is strongly suppressed
near the origin. 

The results obtained in this paper may be useful in any experiment in
which patch potentials could mask forces of different origin, by allowing
for the identification of their contribution, not just their magnitude, but
also their dependence on the geometry of the mirrors. 

\section*{Acknowledgements}
We would like to thank D. Dalvit and R. Decca for useful discussions that stimulated this research.
C.D.F. thanks CONICET, ANPCyT and UNCuyo for financial support. The work of
F.D.M. and F.C.L was supported by UBA, CONICET and ANPCyT.


\appendix
\section{ Expansion of $G$ in powers of $\eta$} \label{A}
Let us present here a sketch of the approach we have used in order to
expand $G$ in powers of $\eta$. To that end we first recall that, in
infinite unbounded space, an electrostatic potential $\varphi({\mathbf
x})$, which is solution to Poisson's equation $\nabla^2 \varphi ({\mathbf
x}) = - \rho ({\mathbf x})$, must be a minimum of the functional
\begin{equation} 
{\mathcal F}[\varphi] \;=\; \int d^3{\mathbf x} \big[
\frac{1}{2} (\nabla \varphi)^2({\mathbf x}) 
- \rho({\mathbf x}) \varphi({\mathbf x}) \big] \;.  
\end{equation}
  Indeed, Poisson's equation is tantamount here to the Euler-Lagrange
equation corresponding to the extremal of the functional.

To consider an spatial region which is limited by the two surfaces $L$ and $R$, where
it satisfies Dirichlet boundary conditions, we may introduce two Lagrange
multipliers $\lambda_L$ and $\lambda_R$,  and construct an augmented functional,
\begin{eqnarray} 
{\mathcal F}[\varphi] &=& \int d^3{\mathbf x} \big[
\frac{1}{2} (\nabla \varphi)^2({\mathbf x}) 
- \rho({\mathbf x}) \varphi({\mathbf x})  \big] \nonumber\\
&+& \int d^2{\mathbf x_\parallel} \left[ 
\lambda_L({\mathbf x_\parallel}) \, \varphi({\mathbf x_\parallel},0)  
+  \lambda_R({\mathbf x_\parallel}) \, \varphi({\mathbf x_\parallel},
\psi({\mathbf x_\parallel}))  \right] \;.  
\end{eqnarray}
Thus, varying the field and the Lagrange multipliers independently, one finds the equation:
\begin{equation}
-\nabla^2 \varphi({\mathbf x}) - \rho({\mathbf x}) \,+\,
\lambda_L({\mathbf x_\parallel}) \, \delta(x_3)
+\lambda_R({\mathbf x_\parallel}) \, \delta(x_3-{\mathbf x_\parallel})
\;=\;0 \;,
\end{equation}
plus
\begin{equation}
\varphi({\mathbf x_\parallel},0) = 0 \;\;,\;\;\;\;
\varphi({\mathbf x_\parallel},\psi({\mathbf x_\parallel})) = 0 \;\;.
\end{equation}
We then recall that $\varphi$ becomes the Dirichlet Green's function: $\varphi \to G({\mathbf
x},{\mathbf y})$ when $\rho({\mathbf x}) \to \delta({\mathbf x} - {\mathbf
y})$. Naturally, Lagrange's multipliers become also functions of the source
point ${\mathbf y}$.

Thus,
\begin{equation}
-\nabla^2_{\mathbf x} G({\mathbf x},{\mathbf y})\;=\; \delta({\mathbf x} - {\mathbf y}) \,
- \, \lambda_L({\mathbf x_\parallel},{\mathbf y}) \, \delta(x_3)
-\lambda_R({\mathbf x_\parallel},{\mathbf y}) \, \delta(x_3-{\mathbf x_\parallel})
\;=\;0 \;.
\end{equation}
Acting with the inverse of the Laplacian on the previous equation, we see
that:
\begin{equation}\label{eq:imp}
G({\mathbf x},{\mathbf y})\;=\; G_0({\mathbf x},{\mathbf y})
- \, \int d^2{\mathbf x'_\parallel} \big[ G_0({\mathbf x};{\mathbf
  x'_\parallel},0) \lambda_L({\mathbf x'_\parallel},{\mathbf y}) 
+ G_0({\mathbf x};{\mathbf
  x'_\parallel},\psi({\mathbf x'_\parallel})) 
\lambda_R({\mathbf x'_\parallel},{\mathbf y}) \big]\;,
\end{equation}
where $G_0$ is the Green's function in the absence of boundaries:
\begin{equation} 
G_0({\mathbf x},{\mathbf y})\,=\, \int \frac{d^3{\mathbf k}}{(2\pi)^3}
\frac{e^{i {\mathbf k}\cdot ({\mathbf x}-{\mathbf y})}}{{\mathbf k}^2} \;.
\end{equation} 
Evaluating (\ref{eq:imp}) for points ${\mathbf x}$ belonging to either $L$
or $R$, we obtain two equations which can be used to determine the Lagrange multipliers:
\begin{equation}
\int d^2{\mathbf x'_\parallel} \big[ G_0({\mathbf x_\parallel},0;{\mathbf
x'_\parallel},0) \lambda_L({\mathbf x'_\parallel},{\mathbf y}) 
+ G_0({\mathbf x_\parallel},0;{\mathbf x'_\parallel},\psi({\mathbf x'_\parallel})) 
\lambda_R({\mathbf x'_\parallel},{\mathbf y}) \big]
= G_0({\mathbf x_\parallel},0;{\mathbf y})
\end{equation}
and
$$
\int d^2{\mathbf x'_\parallel} \big[ G_0({\mathbf x_\parallel},\psi({\mathbf x_\parallel});{\mathbf
  x'_\parallel},0) \lambda_L({\mathbf x'_\parallel},{\mathbf y}) 
+ G_0({\mathbf x_\parallel},\psi({\mathbf x_\parallel});{\mathbf
  x'_\parallel},\psi({\mathbf x'_\parallel})) \lambda_R({\mathbf x'_\parallel},{\mathbf y}) \big]
$$
\begin{equation}
= G_0({\mathbf x_\parallel},\psi({\mathbf x_\parallel});{\mathbf y}) \;.
\end{equation}
Thus, introducing an index $A$ which may assume the values $L$ and $R$, and
two functions $\zeta_L = 0$, $\zeta_R({\mathbf x_\parallel}) = \psi({\mathbf x_\parallel})$, 
we may write the equations for the Lagrange multipliers as follows:
\begin{equation}
\int d^2{\mathbf x'_\parallel} {\mathbb T}_{AB}({\mathbf
x_\parallel},{\mathbf x'_\parallel}) \lambda_B({\mathbf x'_\parallel},{\mathbf y}) 
= G_0({\mathbf x_\parallel},\zeta_A({\mathbf x_\parallel});{\mathbf y}) \;.
\end{equation}
where:
\begin{equation}
{\mathbb T}_{AB}({\mathbf x_\parallel}, {\mathbf x'_\parallel}) \equiv
G_0({\mathbf x_\parallel}, \zeta_A({\mathbf x_\parallel}); {\mathbf
x'_\parallel}, \zeta_B({\mathbf x'_\parallel})) \;. 
\end{equation} 
 
Thus, the Green's function, may be written as follows:
\begin{eqnarray}
G({\mathbf x}, {\mathbf y}) &=& G_0 ({\mathbf x}, {\mathbf y}) 
\nonumber\\
&-& \int_{{\mathbf x'_\parallel},{\mathbf x''_\parallel}} 
\sum_{A,B} \, G_0 ({\mathbf x}; {\mathbf x'_\parallel},\zeta_A({\mathbf x'_\parallel}))
\Delta_{AB}({\mathbf x'_\parallel}, {\mathbf x''_\parallel})
 G_0 ({\mathbf x''_\parallel}, \zeta_B({\mathbf x''_\parallel});{\mathbf y})
\end{eqnarray}
where we introduced the inverse of ${\mathbb T}$:
\begin{equation}
\Delta_{AB}({\mathbf x'_\parallel}, {\mathbf x''_\parallel}) \;=\;
\big[{\mathbb T}^{-1}\big]_{AB}({\mathbf x'_\parallel}, {\mathbf
x''_\parallel})\;.
\end{equation}
Thus, the expansion for $G$ in powers of $\eta$ will be obtained by
expanding $\Delta$ in powers of $\eta$, which in turn requires to expand
${\mathbb T}$. Note, however, that in the
expressions for the different terms contributing to the energy, the Green's
function is evaluated at points that depend on $\psi$, and that that
function also appears in the normal derivatives. Thus, all of them have to
be expanded in order to obtain the energy. 

We conclude by writing the explicit form of the zeroth order term:
\begin{eqnarray}\label{eq:zeroth}
G^{(0)}({\mathbf x}, {\mathbf y}) &=& G_0 ({\mathbf x}, {\mathbf y}) 
\nonumber\\
&-& \int_{{\mathbf w_\parallel},{\mathbf z_\parallel}} 
\sum_{A,B} \, G_0 ({\mathbf x}; {\mathbf w_\parallel},\zeta_A^{(0)}({\mathbf w_\parallel}))
\Delta^{(0)}_{AB}({\mathbf w_\parallel}, {\mathbf z_\parallel})
 G_0 ({\mathbf z_\parallel}, \zeta^{(0)}_B({\mathbf z_\parallel});{\mathbf y})
\end{eqnarray}
where:
\begin{equation}
\Delta^{(0)}_{AB}({\mathbf w_\parallel}, {\mathbf z_\parallel})\;=\;
\int \frac{d^2{\mathbf k_\parallel}}{(2\pi)^2}\;
\widetilde{\Delta}^{(0)}_{AB}({\mathbf k_\parallel}) e^{i {\mathbf
k_\parallel}\cdot ({\mathbf w_\parallel}-{\mathbf z_\parallel})} \;,
\end{equation}
with
\begin{equation}
[\widetilde{\Delta}^{(0)}]({\mathbf k_\parallel})\,=\,
\frac{2 |{\mathbf k_\parallel}|}{ 1 - e^{-2 |{\mathbf k_\parallel}| a}}
\left(
\begin{array}{cc}
1 & - e^{-|{\mathbf k_\parallel}| a}\\
- e^{-|{\mathbf k_\parallel}| a} & 1
\end{array}
\right)
\end{equation}

\end{document}